\documentclass[aps,twocolumn,groupedaddress,nofootinbib,showkeys,showpacs,amsfonts,eqsecnum]{revtex4}

\usepackage{bm}
\usepackage{amsmath}

\newcommand{\R}{{\bm {R}}}
\newcommand{\Rc}{{\cal {R}}}
\newcommand{\tr}{{\textrm {tr}}}
\newcommand{\Tr}{{\textrm {Tr}}}

\begin{document}

\title{Derivative expansion of the heat kernel in curved space}

\author{L.L. Salcedo}

\affiliation{ Departamento de F{\'\i}sica At\'omica, Molecular y
Nuclear, Universidad de Granada, E-18071 Granada, Spain }

\date{\today}

\begin{abstract}
The heat kernel in curved space-time is computed to fourth order in a
strict expansion in the number of covariant derivatives. The
computation is made for arbitrary non abelian gauge and scalar fields
and for the Riemann connection in the coordinate sector. The 
expressions obtained hold for arbitrary tensor representations of the matter
field.  Complete results are presented for the diagonal matrix
elements and for the trace of the heat kernel operator.  In addition,
Chan's formula is extended to curved space-time. As a byproduct, the
bosonic effective action is also obtained to fourth order.
\end{abstract}



\maketitle

\section{Introduction}
\label{sec:1}

The heat kernel operator, the exponential of the Klein-Gordon
operator, is a useful tool in quantum field theory since it is
ultraviolet finite, one-valued and gauge covariant and allows to
obtain the propagator and the effective action
\cite{Hadamard:1952bk,Fock:1937dy,Schwinger:1951nm,Dewitt:1975ys}. Quite
different applications of the heat kernel (such as spectral densities,
index theorems, $\zeta$-function, quantum anomalies, chiral gauge
theories, effective theories of QCD, Casimir effect, black hole
entropies, membranes, etc.)  are illustrated in
\cite{Gilkey:1975iq,Atiyah:1973ad,Hawking:1977ja,%
Fujikawa:1980eg,Ball:1989xg,Bijnens:1996ww,Megias:2005fj,%
Bordag:2001qi,Callan:1994py,Belani:2006wx}.
The exact evaluation of the heat kernel is not possible in general
(see however, \cite{Bytsenko:1996bc,Camporesi:1990wm} and
\cite{Avramidi:1994zp,Avramidi:1995ik,Avramidi:1997jy} for particular
cases). Nevertheless the series expansion in powers of the proper time
is available with computable coefficients, the so-called
Hadamard-Minakshisundaram-DeWitt-Seeley (HMDS) or heat kernel
coefficients \cite{Dewitt:1975ys,Seeley:1967ea}. These coefficients
have been computed with different techniques in several setups and to
dimension ten \cite{Bel'kov:1996tn,vandeVen:1998pf,Moss:1999wq,%
Fliegner:1998rk,Avramidi:1991je,Gusynin:1989ky,Elizalde:1994bk}. For
reviews on spectral geometry and field theory on curved space see
\cite{DeWitt:1965bk,Gilkey:1984bk,Birrel:1982bk,Fulling:1989bk,%
Wald:1994bk,Avramidi:2000bk,Vassilevich:2003xt,Kirsten:2002bk,%
Esposito:2002nc}. For an alternative approach based on covariant
perturbation theory see \cite{Barvinsky:1987uw}. Extensions to finite
temperature can be found in
\cite{Megias:2002vr,Megias:2003ui}. Expansions around non c-number
mass terms are given in
\cite{Osipov:2001bj1,Osipov:2001bj,Salcedo:2001qp}.  The extension to
non commutative quantum field theory has been presented in
\cite{Vassilevich:2003yz}.

In \cite{Salcedo:2004yh} we investigated a different expansion for the
diagonal matrix elements of the heat kernel operator where the terms
are classified by their number of covariant derivatives. This can be
regarded as a resummation of the HMDS coefficients to all orders in
the (non abelian) mass term, which needs not be small. There, explicit
results were presented for boundaryless compact flat manifolds to four
derivatives for the diagonal matrix elements and to six derivatives
for the trace of the heat kernel operator, for Klein-Gordon operators
with arbitrary non abelian gauge and scalar external fields.  In the
present work we extend those results to the case of curved manifolds
endowed with the Levi-Civita connection in the coordinate sector.

The scope and ideas involved as well as some definitions are given in
Section \ref{sec:2}. In particular, we avoid the standard approach of
reducing tensor fields (e.g., the photon field) to scalars by
introduction of a tetrad \cite{Vassilevich:2003xt}. That approach
allows to use the heat kernel formulas derived for coordinate scalars
but at the price of complicating the internal space (due to the new
tetrad index) with its corresponding modified connection. Instead, we
provide formulas which are equally simple regardless of the tensor
representation of the fields (on which the Klein-Gordon operator is
acting) using only the original gauge connection in the sector of
internal indices and the Levi-Civita connection in the coordinate
sector.

In Section \ref{sec:3} we discuss the shortcomings of Chan's approach
in the curved case and develop our own approach based on the use of
covariant symbols \cite{Pletnev:1998yu,Salcedo:2006pv}.  The covariant
symbols are in fact multiplicative operators (with respect to $x$)
which define a faithful representation of the algebra of
pseudodifferential operators.  This technique is similar to that of
symbols for pseudodifferential operators
\cite{Seeley:1967ea,Eguchi:1980jx,Nepomechie:1984wt,Salcedo:1996qy}
and shares with it the feature of providing diagonal matrix elements
of generic operators, not necessarily related to the heat kernel,
since it is not based on recurrence relations. However there is an
important difference between both techniques: in the standard method
of symbols covariance is not manifest prior to momentum integration
whereas with covariant symbols covariance (under both gauge and
coordinate transformations) is manifest at every step of the
calculation. In that section results are provided for the diagonal
matrix elements to four derivatives. These results are presented in
compact form using the technique of labeled operators, also used in
\cite{Salcedo:2004yh}. In this notation the mass term carries a label
indicating its position in the expression; in this way it becomes
effectively a c-number and momentum integrals can be carried out
explicitly.

Explicit formulas for the derivative expansion of the trace of the
heat kernel operator are presented in Section \ref{sec:4}.  This is of
interest in the computation of the effective action. It it is noted
that many, in fact most, of the new terms introduced by the curvature
can be eliminated by a suitable redefinition of the mass term in the
Klein-Gordon operator.

In Section \ref{sec:5} we obtain the expressions for the diagonal
matrix elements and trace of the heat kernel in the form first derived
by Chan \cite{Chan:1986jq} for the effective action to four
derivatives and extended in \cite{Caro:1993fs} to six
derivatives. This is done following a rather indirect path since
Chan's derivation is not used. Instead a Chan's form is proposed and
the coefficients are adjusted to reproduce the results previously
derived using covariant symbols. The question of whether Chan's
elegant method can be used in curved manifolds is left open. In any
case, as expected from the experience in the flat case, Chan's form is
truly much more compact. Drawbacks are that the expressions are less
explicit because one parametric integral is left undone, and for the
same reason there is a by parts integration ambiguity in the formulas.

Section \ref{sec:6} is devoted to computing the bosonic effective
action in an explicit way, with the help of labeled operators. The
necessary momentum integrals can be done with dimensional
regularization. They are given in \cite{Salcedo:2000hp} using the
$\overline{\text{MS}}$ scheme.

Somewhat more explicit details on the calculation with covariant
symbols are given in Appendix \ref{app:A}. Trace cyclic property and
integration by parts identities are derived in Appendix \ref{app:B}.

\section{Derivative expansion of the heat kernel}
\label{sec:2}

We assume a compact boundaryless Riemannian manifold of dimension $d$
and Euclidean metric $g_{\mu\nu}(x)$ to represent the space-time upon
Wick rotation.  Further, for the world sector indices\footnote{ In
this work we will use the label {\em world} interchangeably with {\em
coordinate} or {\em space-time} in expressions like ``world tensor'',
``world index'', etc, to refer to properties tied to indices
$\mu,\nu,\ldots$, associated to natural bases, $\partial/\partial
x^\mu$, of the tangent space of the Riemannian manifold.}  we take the
Levi-Civita connection (i.e., torsionless and metric preserving). The
Klein-Gordon operator will act on wavefunctions $\psi(x)$ (matter
fields) defined on the space-time manifold.  The wavefunctions are
vectors with respect to some representation of a certain gauge group.
Without loss of generality we will assume that $\psi$ carries a single
gauge (or internal) index. $\psi$ is also allowed to carry world
indices, that is, we do not assume $\psi$ to be a world
scalar.\footnote{\label{foot:2} Of course one can choose to transform
such world indices into internal or gauge indices using a tetrad
field. This is the standard approach \cite{Vassilevich:2003xt}.  Our
formulas hold whether this choice is made or not.} Following for
instance \cite{Vassilevich:2003xt}, we use a single covariant
derivative $\nabla_\mu$ which acts on all indices with the appropriate
connection, $\nabla=\partial+\Gamma+\omega$.
$\Gamma^\lambda_{\mu\nu}$ is the connection on world indices and
$\omega_\mu(x)$ the connection on gauge indices, a matrix in internal
space.  Our convention for the Riemann tensor is such that, if
$\psi$ is a world vector,
\begin{equation}
[\nabla_\mu,\nabla_\nu]\psi_\lambda=
\Omega_{\mu\nu}\psi_\lambda
+R_{\mu\nu\lambda\sigma}\psi_\sigma
\label{eq:2.1}
\end{equation}
where
\begin{equation}
\Omega_{\mu\nu} =\partial_\mu\omega_\nu-\partial_\nu\omega_\mu
+[\omega_\mu,\omega_\nu]
\end{equation}
is the gauge field strength tensor. In addition, for the Ricci tensor
and scalar curvature
\begin{equation}
\Rc_{\mu\nu}= R_{\lambda\mu\lambda\nu}\,,
\quad
\R=\Rc_{\mu\mu} \,.
\end{equation}
In the previous formulas we use the same notation for contravariant
and covariant indices. We will follow this convention throughout
unless an ambiguity arises.

The Klein-Gordon operator is of  the form
\begin{equation}
K=g^{\mu\nu}\nabla_\mu\nabla_\nu+X \,.
\end{equation}
$X(x)$ is a scalar field with respect to world indices and a matrix
with respect the internal space.  As is well-known the heat kernel
operator $e^{\tau K}$ is ultraviolet finite for ${\rm Re\,}(\tau)>0$,
and its matrix elements admit an asymptotic expansion in powers of
$\tau$. For the diagonal matrix elements
\begin{equation}
\langle x|e^{\tau K}|x\rangle \asymp
\frac{1}{(4\pi\tau)^{d/2}}\sum_{n=0}^\infty \tau^n a_n(x) \,.
\label{eq:2.5}
\end{equation}
The coefficients $a_n$ are still operators with respect to gauge and
world indices since the brackets $\langle x|$, $|x\rangle$ refer only
to $x$-space. To lowest orders, the well-known result is
\begin{eqnarray}
a_0 &=& 1, 
\nonumber\\
a_1 &=& X+\frac{1}{6}\R,
\nonumber\\
a_2 &=& \frac{1}{2}X^2
+\frac{1}{6}X_{\mu\mu}
+\frac{1}{12}Z_{\mu\nu}^2 
+\frac{1}{6}\R X
+\frac{1}{30}\R_{\mu\mu}
\nonumber \\ &&
+\frac{1}{72}\R^2
-\frac{1}{180}\Rc_{\mu\nu}^2
+\frac{1}{180}R_{\mu\nu\alpha\beta}^2
\,.
\end{eqnarray}
In these formulas (and hereafter) we use the following notational
convention: the covariant derivative of an object is represented by
adding a world index to it and further covariant derivatives add
further indices to the left.  So for instance,
\begin{eqnarray}
X_\mu &=& [\nabla_\mu,X],
\quad 
X_{\mu\nu}=[\nabla_\mu,[\nabla_\nu,X]],
\nonumber \\ 
\Rc_{\alpha\mu\nu} &=& \nabla_\alpha\Rc_{\mu\nu} \,.
\end{eqnarray}

We have also introduced the quantity
\begin{equation}
Z_{\mu\nu}:=[\nabla_\mu,\nabla_\nu] \,.
\end{equation}
Let us emphasize that $Z_{\mu\nu}$ is different from
$\Omega_{\mu\nu}$.  The operators $Z_{\mu\nu}$ and $\Omega_{\mu\nu}$
are both multiplicative with respect to $x$-space and matrices in
internal space, however, unlike $\Omega_{\mu\nu}$, $Z_{\mu\nu}$ acts
also on world indices (cf.  (\ref{eq:2.1})).  This implies, for
instance, that $Z_{\mu\nu}$ and $\Rc_{\alpha\beta}$ do not commute:
\begin{equation}
[Z_{\mu\nu},\Rc_{\alpha\beta}] =
R_{\mu\nu\alpha\sigma}\Rc_{\sigma\beta} +
R_{\mu\nu\beta\sigma}\Rc_{\alpha\sigma} 
\,.
\end{equation}
Likewise, $[Z_{\mu\nu},X_\lambda]= [\Omega_{\mu\nu},X_\lambda]
+R_{\mu\nu\lambda\sigma}X_\sigma$, etc.  The derivative of
$Z_{\mu\nu}$ will also be needed,\footnote{This is an exception to the
index convention. The second term in the definition of
$Z_{\alpha\mu\nu}$ is required to make it a multiplicative operator
with respect to $x$. E.g.
$$
[Z_{\alpha\mu\nu},X_\lambda]= [\Omega_{\alpha\mu\nu},X_\lambda] 
+R_{\alpha\mu\nu\lambda\sigma}X_\sigma
\,.
$$
Similar terms appear in higher derivatives, $Z_{\alpha\beta\mu\nu}$,
etc, \cite{Salcedo:2006pv}.}
\begin{equation}
Z_{\alpha\mu\nu} := [\nabla_\alpha,Z_{\mu\nu}]
-\frac{1}{2}\{\nabla_\lambda,R_{\lambda\alpha\mu\nu}\}
\,.
\end{equation}

In our formulas we will use $Z_{\mu\nu}$ rather than
$\Omega_{\mu\nu}$. Nevertheless, if desired one can always move the
$Z$'s to the right using their commutation properties and apply them
to the wavefunction $\psi$ to produce $\Omega$'s and $R$'s.  (Of
course, the result so obtained will depend on the concrete tensor
representation of the wavefunction.)  For instance, the term
$Z_{\mu\nu}^2/12$ in $a_2$ is equivalent to more standard
$\Omega_{\mu\nu}^2/12$ when the wavefunction happens to be a world
scalar. The advantage of using $Z_{\mu_1\cdots\mu_n}$ is that the
expressions take the same form regardless of the world-tensor
representation of $\psi$ and yet the original connections $\Gamma$ and
$\omega$ are used. (See footnote \ref{foot:2}.)  Fuller details can be found in
\cite{Salcedo:2006pv}.

By dimensional counting it is clear that the standard heat kernel
expansion (\ref{eq:2.5}) is an expansion with coefficients ordered by
their mass dimension: if $g_{\mu\nu}$ carries dimension 0,
$\nabla_\mu$ carries dimension 1, and $X$ carries dimension 2, the
coefficient $a_n$ has dimension $2n$. In what follows we will set
$\tau=1$ since dimensional counting allows to restore the parameter
$\tau$ at any time if needed.

In this work we consider a different classification of terms in the
heat kernel, namely, operators are classified by the number of
covariant derivatives they carry, so $\nabla_\mu$ counts as order 1
while $X$ and $g_{\mu\nu}$ count as order cero. Thus, e.g., $X X_\mu$
is of first order, $R_{\mu\nu\alpha\beta}$ and $Z_{\mu\nu}$ of second
order, $Z_{\alpha\mu\nu}$ of third order, and so on. In this expansion
\begin{equation}
\langle x|e^K|x\rangle \asymp
\frac{1}{(4\pi)^{d/2}}\sum_{n=0}^\infty  A_n(x) \,,
\label{eq:2.12}
\end{equation}
where the coefficient $A_n$ collects all operators with $2n$
derivatives and any number of $X$. In turn, the $A_n$ can be
reexpanded using operators classified by their mass dimension,
\begin{eqnarray}
A_0 &=& 1+ X+\frac{1}{2}X^2+\cdots, 
\nonumber\\
A_1 &=& 
\frac{1}{6}\R
+\frac{1}{6}X_{\mu\mu}
+\frac{1}{6}\R X
+\cdots,
\nonumber\\
A_2 &=& 
\frac{1}{12}Z_{\mu\nu}^2 
+\frac{1}{30}\R_{\mu\mu}
+\frac{1}{72}\R^2
-\frac{1}{180}\Rc_{\mu\nu}^2
\nonumber \\ &&
+\frac{1}{180}R_{\mu\nu\alpha\beta}^2
+\cdots\,.
\end{eqnarray}
(Of course, when $\tau$ is not set to unity these coefficients are
also functions of $\tau$.)

In \cite{Salcedo:2004yh} we presented complete formulas (i.e., valid to all
orders in $X$) for $A_0$, $A_1$ and $A_2$, as well as $B_0$, $B_1$,
$B_2$ and $B_3$ (defined below), for the flat case. Presently we
extend those results (except $B_3$) to curved manifolds.

The key tool to write the result to all orders in $X$ in closed form
is the use of labeled operators. For instance, all terms in $A_1$ of
the type $X^n X_{\mu\mu}X^m$, with $n,m=0,1,\ldots$, can be collected
as
\begin{equation}
I_{2,2}X_{\mu\mu}=
\left(
\frac{e^{X_1} + e^{X_2}}{(X_1 - X_2)^2} - 
  2\frac{e^{X_1} - e^{X_2}}{(X_1 - X_2)^3}
\right)
X_{\mu\mu} \,.
\end{equation}
The label 1 in $X_1$ indicates that the corresponding $X$ should be
placed just before (i.e. to the left of) the fixed operator
$X_{\mu\mu}$, likewise, $X_2$ indicates that that $X$ is to be put
just after (i.e., to the right of) the fixed operator.  Upon a series
expansion in powers of $X_1$ and $X_2$
\begin{eqnarray}
I_{2,2}X_{\mu\mu} &=&
\left(
\frac{1}{6} + \frac{1}{12}X_1 + \frac{1}{12}X_2
\right.
\nonumber \\ &&
\left.
+\frac{1}{40}X_1^2+\frac{1}{40}X_2^2+\frac{1}{30}X_1 X_2
+\cdots
\right)
X_{\mu\mu} 
\nonumber \\
&=&
\frac{1}{6}X_{\mu\mu}  
+ \frac{1}{12}X X_{\mu\mu}   + \frac{1}{12} X_{\mu\mu}  X
\nonumber \\ &&
+\frac{1}{40}X^2 X_{\mu\mu}  
+\frac{1}{40} X_{\mu\mu}  X^2
+\frac{1}{30}X X_{\mu\mu}   X
\nonumber \\ &&
+\cdots
\,.
\end{eqnarray}
Likewise, $X_1^2 X_2 X_3 X_\mu^2$ would stand for $X^2 X_\mu X X_\mu
X$, etc. The labels always refer to the position of $X$'s with respect
to ``fixed operators'' such as $X_{\alpha\mu\nu}$, $Z_{\mu\nu}$,
etc. Because $X$ commutes with $R_{\mu\nu\alpha\beta}$ we will not
need to include the Riemann tensor among the set of fixed
operators. So for instance, $I_{2,2}\R X_{\mu\mu}$ will used to mean
$\R$ multiplied by $I_{2,2} X_{\mu\mu}$.

The important point is that the labeled operators can be treated as
c-numbers, e.g., $X_1X_2=X_2X_1$; the true position of the labeled
operator is given by its label. This is similar to what happens within
normal or chronological orders. Labeled operators were introduced in
\cite{Salcedo:1998sv,Garcia-Recio:2000gt}.

The function $I_{2,2}$ belongs to the family of functions
$I_{r_1,\dots,r_n}$, with arguments $X_1,\ldots,X_n$. They are defined as
\begin{equation}
I_{r_1,r_2,\ldots,r_n} :=
\int_\Gamma\frac{dz}{2\pi i}e^z N_1^{r_1}N_2^{r_2}\cdots N_n^{r_n} \,,
\label{eq:2.15}
\end{equation}
where
\begin{equation}
N=(z-X)^{-1}
\label{eq:2.16}
\end{equation}
and $N_i=(z-X_i)^{-1}$, and $\Gamma$ is a positively oriented closed
simple path on the complex plane enclosing the eigenvalues of $X$ (at
the given point $x$).  These functions enjoy a number of properties
described in \cite{Salcedo:2004yh}. In particular they are entire
functions of the $X_i$ and can be computed by recurrence relations
starting from $I_1=e^{X_1}$. They are not linearly independent;
integration by parts implies the relation
\begin{equation}
I_{r_1,r_2,\ldots,r_n} =\sum_{j=1}^nr_j
I_{r_1,r_2,\ldots,r_j+1,\ldots,r_n} \,.
\label{eq:2.18}
\end{equation}

The trace of the heat kernel operator,
\begin{equation}
\Tr(e^K)= \int d^dx\sqrt{g}\,\tr\langle x|e^K|x\rangle
\,,
\label{eq:2.19}
\end{equation}
is also of great interest in applications, such as the computation of
the effective action (see Section \ref{sec:6}). The trace $\tr$ refers
to gauge and world indices. The trace of the heat kernel admits an
asymptotic expansion with terms classified by their mass dimension:
\begin{equation}
\Tr(e^K) \asymp
\frac{1}{(4\pi)^{d/2}}\sum_{n=0}^\infty  
\int d^dx\sqrt{g}\,\tr(b_n(x)) \,,
\end{equation}
with
\begin{eqnarray}
b_0 &=& 1, 
\nonumber\\
b_1 &=& X+\frac{1}{6}\R,
\nonumber\\
b_2 &=& \frac{1}{2}X^2
+\frac{1}{12}Z_{\mu\nu}^2 
+\frac{1}{6}\R X
\nonumber \\ &&
+\frac{1}{72}\R^2
-\frac{1}{180}\Rc_{\mu\nu}^2
+\frac{1}{180}R_{\mu\nu\alpha\beta}^2
\,.
\end{eqnarray}

Likewise, the terms can be classified by their number of derivatives
\begin{equation}
\Tr(e^K) \asymp
\frac{1}{(4\pi)^{d/2}}\sum_{n=0}^\infty  \int d^dx\sqrt{g}\,\tr(B_n(x)) 
\,.
\end{equation}
Due to the trace cyclic property and integration by parts, there is an
ambiguity in the definition of $B_n(x)$. This allows to bring $B_n(x)$
to a simpler form starting from $A_n(x)$. In turn, functional
derivation allows one to obtain $A_n$ from $B_n$
\begin{equation}
A_n(x)= \frac{\delta}{\delta X(x)}\int d^dx\sqrt{g}\,\tr(B_n(x)) \,.
\label{eq:2.22}
\end{equation}

Before closing this Section, let us comment on the nature of the
derivative expansion in the present context. [We reinstate $\tau$ for
this discussion.] As is well-known, the standard expansion in powers
of $\tau$ is an asymptotic one, reliable for small $\tau$
only. Intuitively, this is because the coefficients $a_n(x)$ are
local, i.e., they depend on a finite number of derivatives of the
external fields (namely, the metric, the gauge connection and the
scalar field $X$). Consequently, the expansion at a given point $x_0$
is not sensitive to modifications of these external fields taking
place outside a fixed neighborhood of $x_0$. Such modifications affect
the exact matrix element at $x_0$ (and hence its $\tau$ dependence)
but this is not seen by the asymptotic expansion.  By the same token,
the derivative expansion is expected to be asymptotic too, since the
coefficients $A_n(x;\tau)$ are also local. To make mathematical sense
of the asymptotic series it is necessary to write it as a power series
expansion.  The derivative expansion can be viewed as a power series
as follows: for a given point $x_0$ consider a deformation of the
external fields such that within a fixed neighborhood of $x_0$ the
deformation is just a dilatation by a parameter $\lambda$
\begin{eqnarray}
X(x)\mapsto X(x_\lambda), 
\nonumber \\
g_{\mu\nu}(x)\mapsto g_{\mu\nu}(x_\lambda), 
\nonumber \\
\omega_\mu(x)\mapsto \lambda\omega_\mu(x_\lambda),
\nonumber \\
x_\lambda^\mu=x^\mu_0+\lambda(x^\mu-x^\mu_0) \,.
\end{eqnarray}
The deformation is smoothly continued outside of the
neighborhood. This produces a family of Klein-Gordon operators
$K_\lambda$.\footnote{The dilatation depends not only on $x_0$ but
also on the coordinate system used. Eq. (\ref{eq:2.24}) is coordinate
independent.} By construction the parameter $\lambda$ counts the
number of derivatives, that is,
\begin{equation}
A_n(x_0;\tau,\lambda) = \lambda^{2n} A_n(x_0;\tau) \,.
\label{eq:2.24}
\end{equation}
Therefore the derivative expansion can be read off from an expansion
of $\langle x_0|e^{\tau K_\lambda}|x_0\rangle$.  This construction
suggests that, although the series in $\lambda$ is asymptotic, the
coefficients $A_n(x;\tau)$ themselves are well defined quantities, as
is also the case for the coefficients $a_n(x)$. This point is to be
settled by a rigorous mathematical approach.  Note that this
definition of the coefficients (from an expansion in $\lambda$) is not
identical to the alternative definition
\begin{equation}
A_n(x;\tau)=\sum_{m\ge n}a_{m,n}(x)\tau^{m-n}
\end{equation}
(using $a_{m,n}(x)$ to denote the terms of $a_m(x)$ with $2n$
derivatives).  We expect both definitions to coincide. This
expectation relies on (i) the fact that the functions
$I_{r_1,r_2,\ldots,r_n}$ are entire functions of $X_i$ and so of
$\tau$, thus their expansion in $\tau$ is absolutely convergent, and
(ii) the obvious reason for the series in $\tau$ to be asymptotic does
not apply here since $A_n(x;\tau)$ is itself local. In summary, we
expect the derivative expansion to be valid for finite (not small)
$\tau$, provided $\lambda$ is sufficiently small.

\section{Diagonal coefficients}
\label{sec:3}

In \cite{Salcedo:2004yh}  the calculation of $A_n$  in flat space-time
was based on that of $B_n$.  In turn $B_n$ was adapted from the result
of  Chan   \cite{Chan:1986jq}  for   the  effective  action   to  four
derivatives   and   its  extension   in   \cite{Caro:1993fs}  to   six
derivatives. Unfortunately, it is not obvious how the elegant approach
of \cite{Chan:1986jq} is to be extended  to the curved case. As we may
recall, in that approach a symbol  method ($D_\mu \to D_\mu + p_\mu $)
is  applied,  to  $\Tr(\log(K))$.  The  expression  is  then  formally
expanded in the number of  derivatives and brought to a canonical form
by using  integration by  parts (with respect  to $p_\mu$) as  well as
formal cyclic  property.  This canonical form is  not manifestly gauge
invariant  but it  allows  to unambiguously  identify the  originating
gauge  invariant  expression  from  which  it comes,  because  no  two
different  gauge invariant  expressions  may have  the same  canonical
form. The  trouble is that  a similar statement  does not hold  in the
curved case. For instance,
\begin{equation}
 \Tr(Z_{\mu\nu}[p_\mu,p_\nu N^2])=0
\end{equation}
(with $N=(p^2-X)^{-1}$), since $p_\mu$ and $X$ commute. However, 
under formal cyclic property the same expression would be equivalent to
\begin{eqnarray}
 \Tr(p_\nu N^2[Z_{\mu\nu},p_\mu]) &=& 
\Tr(p_\nu N^2 R_{\mu\nu\mu\sigma}p_\sigma)
\nonumber \\ 
&=& \Tr(p_\mu p_\nu\Rc_{\mu\nu}N^2) \,.
\end{eqnarray}
This is equivalent to $-\Tr(p^2 \R N^2)/d$ and does not vanish.

Since Chan's approach is not available, we will compute $A_n$ from
scratch and then obtain $B_n$ from it. The starting point is the
representation
\begin{equation}
e^K=\int_\Gamma\frac{dz}{2\pi i}\frac{e^z}{z-\nabla^2-X}
\end{equation}
where we can apply the method of covariant symbols
\cite{Pletnev:1998yu,Salcedo:2006pv}. This is the second key
ingredient of our approach. This gives
\begin{equation}
\langle x|e^K|x\rangle 
=\int_\Gamma\frac{dz}{2\pi i}
\frac{1}{\sqrt{g}} \int \frac{d^dp}{(2\pi)^d}
\frac{e^z}{z-{\overline\nabla}^2-\overline{X}}
\,.
\label{eq:3.4}
\end{equation}
The covariant symbols are designed for computing diagonal matrix
elements of general operators, no necessarily the heat kernel. The
covariant symbols were introduced by Pletnev and Banin in
\cite{Pletnev:1998yu} for the gauge connection and extended to curved
space-time in \cite{Salcedo:2006pv} where they have been computed to
four derivatives. Explicitly, to two derivatives
\begin{eqnarray}
{\overline{X}} &=& 
X - X_\alpha\,\partial^\alpha
+\frac{1}{2!}X_{\alpha\beta}\,\partial_\alpha\partial_\beta
+\cdots \,,
\nonumber
\\
\overline{\nabla}_\mu^2
 &=&
p_\mu^2
+\frac{1}{6}\R
+ Z_{\alpha \beta} \, p_\alpha \partial_\beta
-\frac{1}{3}\Rc_{\alpha \beta} \, p_\alpha \partial_\beta
\nonumber \\ &&
+\frac{1}{3} R_{\alpha \mu \beta \nu} \,
p_\alpha p_\beta \partial_\mu\partial_\nu
+\cdots \,,
\label{eq:3.5}
\end{eqnarray}
where $\partial_\mu=\partial/\partial p_\mu$ and the dots refer to
terms with three derivatives or more.  Note that, unlike ordinary
symbols, these operators are multiplicative and covariant already
before momentum integration. The calculation is straightforward along
the lines of the examples presented in \cite{Salcedo:2006pv}.  Details
are provided in Appendix \ref{app:A}.  One obtains
\begin{equation}
A_0= I_1=e^X
\label{eq:3.6}
\end{equation}
to zeroth order and
\begin{eqnarray}
A_1 &=& 
(I_{1,2}-2I_{1,3})X_{\mu\mu} + (2I_{1,1,2}-4I_{1,1,3}-2I_{1,2,2})X_\mu^2
\nonumber \\ &&
+\frac{1}{3} I_3\R
\label{eq:3.7}
\end{eqnarray}
to second order. Using the identities (\ref{eq:2.18}), this can
be rewritten as
\begin{equation}
A_1= 
I_{2,2} X_{\mu\mu} + 2I_{2,1,2} X_\mu^2 
+\frac{1}{6} I_1\R
\,.
\end{equation}
Let us emphasize that the functions $f(X_1,X_2)$ multiplying
$X_{\mu\mu}$ and $f(X_1,X_2,X_3)$ multiplying
$X_\mu^2$, are well defined and unambiguous. The only
ambiguity enters in how they are written in terms of the overcomplete
basis $I_{r_1,\ldots,r_n}$.

The two terms without $\R$ are identical to those of the flat case in
\cite{Salcedo:2004yh}, only with the ``minimal coupling'' replacement
$D=\partial+\omega \to \nabla =\partial+\Gamma+\omega$. These are then
minimal terms required by covariance (under general coordinate
transformations). The term with $\R$ is non minimal; covariant but not
required by covariance.  In general we will obtain coefficients of the
form
\begin{equation}
A_n= A_n^m + A_n^R
\,.
\end{equation}
The terms in $A_n^R$ are those which contain explicitly the Riemann
tensor and so vanish in the flat space case. On the other hand, the
minimal terms $A_n^m$ can be reconstructed by minimal coupling from the
flat space expressions. Let us warn, however, that this separation is
not an unambiguous one: in general $A_n^m$ will depend of the concrete
$A_n$ of flat space used to apply the minimal coupling. This is
because reordering of world indices in the flat space expression may
introduce Riemann tensors in the curved case.\footnote{For instance,
$$
Z_{\alpha\beta\mu\nu}=
Z_{\beta\alpha\mu\nu} 
+[Z_{\alpha\beta},Z_{\mu\nu}]
-\frac{1}{2}\{Z_{\alpha\lambda}, R_{\lambda\beta\mu\nu}\}
+\frac{1}{2}\{Z_{\beta\lambda}, R_{\lambda\alpha\mu\nu}\} \,.
$$
}

The calculation of $A_2$ gives 
\begin{widetext}
\begin{eqnarray}
A_2^m &=&
  2 I_{2,1,2} \,Z_{\mu \nu} Z_{\mu \nu} 
\nonumber \\ && 
+ ( 2 I_{2,2,2} - 4 I_{3,0,3} + 4 I_{2,1,3} )\,Z_{\mu \mu \nu} X_{\nu} 
\nonumber \\ && 
+ ( 2 I_{2,2,2} - 4 I_{3,0,3} + 4 I_{3,1,2} )\,X_{\mu} Z_{\nu \mu \nu} 
\nonumber \\ && 
+   ( 4 I_{2,2,1,2} - 16 I_{3,0,1,3} - 8 I_{3,0,2,2} +  8 I_{3,1,1,2}
)\,
     X_{\mu} Z_{\mu \nu} X_{\nu} 
\nonumber \\ && 
+ (16 I_{3,0,1,3} - 8 I_{2,1,1,3} - 4 I_{2,1,2,2} + 
      8 I_{3,0,2,2})\, Z_{\mu \nu} X_{\mu} X_{\nu} 
\nonumber \\ && 
+ (16 I_{3,1,0,3} - 8 I_{3,1,1,2} - 4 I_{2,2,1,2} + 
      8 I_{2,2,0,3})\, X_{\mu} X_{\nu} Z_{\mu \nu} 
\nonumber \\ && 
 + 2 I_{3,3} \,X_{\mu \mu \nu \nu} 
\nonumber \\ && 
+ ( 2 I_{2,3,2} + 4 I_{3,1,3} + 2 I_{2,2,3}+ 2 I_{3,2,2} )\,
     X_{\mu \mu}  X_{\nu \nu} 
\nonumber \\  &&
+ 8 I_{3,1,3} \,X_{\mu \nu} X_{\mu \nu} 
\nonumber \\  &&
+  (8 I_{3,1,3} + 4 I_{3,2,2})\, X_{\mu \nu \nu} X_{\mu} 
\nonumber \\  &&
+  (8 I_{3,1,3} + 4 I_{2,2,3})\,X_{\mu} X_{\mu \nu \nu}
\nonumber \\ && 
+ (  4 I_{2,2,2,2} +  16 I_{3,1,1,3} + 8 I_{2,2,1,3} + 8 I_{3,1,2,2})\, 
X_{\mu} X_{\mu \nu} X_{\nu} 
\nonumber \\ && 
+ (  2 I_{2,2,2,2} + 8 I_{3,1,1,3}
   + 4 I_{2,2,1,3}+4 I_{3,1,2,2} )\, X_{\mu} X_{\nu \nu} X_{\mu} 
\nonumber \\ && 
+ (  2 I_{2,2,2,2} + 8 I_{3,1,1,3} + 4 I_{2,2,1,3} 
+ 4 I_{2,3,1,2} + 4 I_{3,1,2,2} + 4 I_{3,2,1,2})\,
     X_{\mu \mu} X_{\nu} X_{\nu} 
\nonumber \\ && 
+ (  2 I_{2,2,2,2} + 8 I_{3,1,1,3} + 4 I_{2,2,1,3}
+ 4 I_{2,1,3,2} + 4 I_{3,1,2,2}   + 4 I_{2,1,2,3})\,
    X_{\mu} X_{\mu}  X_{\nu \nu} 
\nonumber \\ && 
+ (16 I_{3,1,1,3} + 8 I_{3,1,2,2})\,X_{\mu \nu} X_{\mu} X_{\nu} 
\nonumber \\ && 
+ (16 I_{3,1,1,3} + 8 I_{2,2,1,3})\,X_{\mu} X_{\nu} X_{\nu \mu} 
\nonumber \\ &&
  + (4 I_{2,2,1,2,2}  +  16 I_{3,1,1,1,3} 
 + 8 I_{2,1,2,1,3} + 8 I_{3,1,2,1,2} 
+ 4 I_{2,2,2,1,2} + 4 I_{2,1,2,2,2}
\nonumber \\ &&
\qquad
+ 8 I_{2,1,3,1,2} 
+ 8 I_{2,2,1,1,3} + 8 I_{3,1,1,2,2} 
)\, X_{\mu} X_{\mu} X_{\nu} X_{\nu} 
\nonumber \\ && 
 + (  4 I_{2,2,1,2,2} + 16 I_{3,1,1,1,3} 
+ 8 I_{2,2,1,1,3} + 8 I_{3,1,1,2,2}
)\, X_{\mu} X_{\nu} X_{\mu} X_{\nu} 
\nonumber \\ && 
 + ( 4 I_{2,2,1,2,2} + 16 I_{3,1,1,1,3}
 + 8 I_{2,2,1,1,3} + 8 I_{3,1,1,2,2}
)\,  X_{\mu} X_{\nu} X_{\nu} X_{\mu} \,.
\label{eq:3.10}
\end{eqnarray}
\begin{eqnarray}
A_2^R &=& 
\frac{1}{30} I_1\R_{\mu\mu}
+\frac{1}{72} I_1\R^2
+\frac{1}{6} I_{2,2}\R X_{\mu\mu}
+\frac{1}{4} I_{2,2}\R_\mu X_\mu
+\frac{4}{3} I_{3,3} \Rc_{\mu\nu} X_{\mu\nu}
+\frac{1}{3} I_{2,1,2} \R X_\mu^2
\nonumber \\ &&
+(\frac{2}{3} I_{2,2,2} + \frac{4}{3} I_{2,3,2} -\frac{16}{3} I_{3,1,3} ) 
  \Rc_{\mu\nu} X_\mu X_\nu
-\frac{1}{180} I_1 \Rc_{\mu\nu}^2
+\frac{1}{180} I_1 R_{\mu\nu\alpha\beta}^2
\,.
\label{eq:3.11}
\end{eqnarray}
\end{widetext}

$A_2^m$ is formally identical to the expression in (4.10) of
\cite{Salcedo:2004yh}. (There mirror symmetry of $A_n$ was exploited
to write $A_2$ in a shorter, less explicit form.) As said before, in
the formula for $A_2^R$ the Riemann tensor is not to be considered as
one of the fixed operators.

The coefficients $A_1$ and $A_2$, for curved space and non abelian
gauge group, with contributions to all orders in $X$ are computed here
for the first time. Upon expansion in powers of $X$ they reproduce the
corresponding terms with up to four derivatives of the standard
expansion $a_n$. In particular the coefficients quoted in
Eqs. (4.26-29) of \cite{Vassilevich:2003xt} are correctly
reproduced. (The comparison with \cite{Vassilevich:2003xt} is achieved
by restricting our results to the case of world scalar wavefunction
and using the trace cyclic property, but not integration by parts.)
Results to all orders in $X$ and up to four derivatives for curved
space are also available from \cite{Gusynin:1990bu} for the so-called
minimal case, i.e., $X(x)$ a c-number and $\omega_\mu=0$, and $\psi$ a
world scalar. Our formulas also check Eq. (22) of
\cite{Gusynin:1990bu}.

\section{Coefficients for the trace}
\label{sec:4}

Application of integration by parts and the trace cyclic property
allows to obtain the simpler coefficients $B_n$. To four derivatives
they take the form
\begin{eqnarray}
B_0 &=& I_1 \,, 
\nonumber \\ 
B_1 &=& -\frac{1}{2} I_{2,2}X_\mu X_\mu
+\frac{1}{6} I_1\R \,, 
\nonumber \\ 
B_2 &=&
(-I_{2,2,2,2}+4I_{3,1,3,1})X_\mu X_\mu X_\nu X_\nu 
\nonumber \\ && 
+ \frac{1}{2}I_{2,2,2,2}X_\mu X_\nu X_\mu X_\nu 
+ 4I_{3,1,3}X_\mu X_\mu X_{\nu\nu} 
\nonumber \\ && 
+ I_{3,3}X_{\mu\mu}X_{\nu\nu} +
2I_{2,2,2}X_\mu X_\nu Z_{\mu\nu} 
+ \frac{1}{2}I_{2,2}Z_{\mu\nu}Z_{\mu\nu} 
\nonumber \\ && 
-\frac{1}{12} I_{2,2} \R X_\mu X_\mu 
- \frac{2}{3} I_{3,3} \Rc_{\mu\nu} X_\mu X_\nu
\nonumber \\ && 
+ \frac{1}{30} I_1\R_{\mu\mu} 
+ \frac{1}{72} I_1\R^2
\nonumber \\ && 
- \frac{1}{180} I_1 \Rc_{\mu\nu}^2 
+ \frac{1}{180} I_1 R_{\mu\nu\alpha\beta}^2 
\,.
\label{eq:4.1}
\end{eqnarray}
[For short we have written $I_{2,2}$ for $I_{2,2,0}$, etc.]  The
simplest way to obtain this result is starting from its Chan's form,
to be discussed in the next section. The validity of integration by
parts for general world representations of the wavefunction space is
shown in the appendix. The cyclic property for multiplicative
operators is also subtle when $Z_{\mu\nu}$ (or more generally
$Z_{\mu_1\cdots\mu_n}$) is present because this operator acts on world
indices. For instance, under trace,
\begin{equation}
 X_\mu Z_{\mu\nu} X_\nu \equiv
  X_\nu  X_\mu Z_{\mu\nu} -\Rc_{\mu\nu} X_\mu X_\nu \,.
\label{eq:4.2}
\end{equation}
The last term would not be present under the standard trace cyclic
property of matrices. The rationale of the extra term is that in
$X_\mu Z_{\mu\nu} X_\nu $, $Z_{\mu\nu}$ acts on the index $\nu$ of
$X_\nu$ (as well as on the subsequent world indices in the
wavefunction) while in $X_\nu X_\mu Z_{\mu\nu}$ it does not; the
missing contribution is added by the extra term.  This equation is
derived in Appendix \ref{app:B}.\footnote{Of course, relations like
(\ref{eq:4.2}) imply that starting from the same $B_n$ of flat space
but written in different ways will in general yield different results
for the minimal coupling part of $B_n$ in curved space.}

It can be verified that $A_n$ and $B_n$ are indeed equivalent inside
$\int d^dx\sqrt{g}\,\tr(~)$ and that (\ref{eq:2.22}) is fulfilled.

To this order one can see that most of the non minimal terms can be
generated by a modified ``minimal coupling'' prescription. Namely,
$\partial+\omega\to\partial+\omega+\Gamma$ and $X\to X^\prime$, with
the new scalar field
\begin{equation}
X^\prime= X +\frac{1}{6}\R +\frac{1}{180}(
\R_{\mu\mu}-\Rc_{\mu\nu}^2+R_{\mu\nu\alpha\beta}^2)
+{\cal O}(\nabla^6) \,.
\end{equation}
The trace of the heat kernel can then be
expanded as
\begin{equation}
\Tr(e^K) \asymp
\frac{1}{(4\pi)^{d/2}}\sum_{n=0}^\infty  
\int d^dx\sqrt{g}\,\tr(B_n^\prime(x)) \,,
\end{equation}
with
\begin{eqnarray}
B_0^\prime &=& I_1^\prime  \,,
\nonumber \\ 
B_1^\prime &=& -\frac{1}{2} I_{2,2}^\prime X^\prime_\mu X^\prime_\mu
 \,,
\nonumber \\
B_2^\prime &=&
(-I_{2,2,2,2}^\prime+4 I_{3,1,3,1}^\prime)
X^\prime_\mu X^\prime_\mu X^\prime_\nu X^\prime_\nu
\nonumber \\ &&
+
\frac{1}{2}I^\prime_{2,2,2,2}
X^\prime_\mu X^\prime_\nu X^\prime_\mu X^\prime_\nu
+
4I^\prime_{3,1,3}X^\prime_\mu X^\prime_\mu X^\prime_{\nu\nu}
\nonumber \\ &&
+
I^\prime_{3,3}X^\prime_{\mu\mu}X^\prime_{\nu\nu}
+
2I^\prime_{2,2,2}X^\prime_\mu X^\prime_\nu Z_{\mu\nu}
+
\frac{1}{2}I^\prime_{2,2}Z_{\mu\nu}Z_{\mu\nu}
\nonumber \\ &&
-\frac{2}{3} I^\prime_{3,3} \Rc_{\mu\nu} X^\prime_\mu X^\prime_\nu
\,.
\end{eqnarray}
($I^\prime_{r_1,\ldots,r_n}$ being defined as in (\ref{eq:2.15}) but
using $X^\prime$ instead of $X$.) There are corresponding coefficients
$A_n^\prime$.  The relations $\int d^dx\sqrt{g}\,\tr(A_n) =\int
d^dx\sqrt{g}\,\tr(B_n)$ and (\ref{eq:2.22}) hold also for the primed
coefficients. The redefinition $X^\prime= X +\frac{1}{6}\R$ is quite
standard in the literature \cite{Bytsenko:1996bc} to eliminate some of
the terms. Note that beyond that the primed expansion is no longer a
strict expansion in the number of covariant derivatives (and
$X^\prime$ will depend on $\tau$ when $\tau$ is restored).

\section{Chan's form of the coefficients}
\label{sec:5}

Let us call the coefficients $A_n$ and $B_n$ just derived the
coefficients in $X$-form, to distinguish them from their Chan's or
$N$-form, to be discussed in this section.

Let us briefly summarize Chan's method in flat space
\cite{Chan:1986jq,Caro:1993fs}.  In this method the results are
obtained in terms of derivatives of $N=(z-X)^{-1}$, instead of
derivatives of $X$, that is, $N_\mu= [\nabla_\mu N]$, $N_{\mu\nu}$,
etc.  Because $z$ appears inside $N_{\mu_1\cdots\mu_n}$ the integral
over $z$ cannot be carried out explicitly and is left undone.
Integration by parts (with respect to $z$) allows to reorder terms so
that in each term of $B_n$ the quantity $N$ (derivated or not) appears
exactly $2n$ times. A virtue of this approach is that for each $B_n$
there is only a limited number of available covariant structures
constructed with $2n$ $N$'s and $2n$ $\nabla$'s, thus the expressions
so obtained are quite compact.  To pass a result given in $N$-form to
$X$-form is, of course, straightforward using the relation $N_\mu= N
X_\mu N$ and its derivatives. As pointed out before, the result in
$X$-form is free from ambiguities.

Because the extension of Chan's method to curved space is not known,
the existence of a Chan's form for the heat kernel coefficients in the
curved case is not obvious.  In principle, undoing the $z$ integrals
in the $I_{r_1,\ldots,r_n}$ and using identities of the type
\begin{eqnarray}
X_\mu &=& N^{-1} N_\mu N^{-1}\,,
\nonumber \\
X_{\mu\nu} &=&
 N^{-1} N_{\mu\nu} N^{-1} 
- N^{-1} N_\mu N^{-1} N_\nu N^{-1}
\nonumber \\ &&
- N^{-1} N_\nu N^{-1} N_\mu N^{-1} \,,
\end{eqnarray}
would allow to bring the coefficients computed in $X$-form to an
almost Chan's form, except that, in general, negative powers of $N$
will be present. Moreover the precise result will be subject to
ambiguities due to integration by parts on $z$ and not algorithm is
available to bring it to a compact form. This method works for $B_0$
and $B_1$.

The procedure that we have followed to obtain an $N$-form for the
coefficient $B_2$ is as follows. In the flat case the $N$-form is
known for $B_2$ and hence for $A_2$ (by functional variation with
respect to $X$), therefore we apply minimal coupling there.  This
already reproduces most of the terms of the known full $A_2$ in
$X$-form, (\ref{eq:3.10}-\ref{eq:3.11}). For the few remaining terms
of $A_2$ it is relatively easy to bring them to a rather compact
$N$-form by hand. This remainder is the functional variation of the
non minimal remainder in $B_2$, $B_2^R$, not yet determined. To obtain
$B_2^R$ we simply write down the most general terms having four
derivatives, at least one Riemann tensor and no more than four $N$'s
and with arbitrary numerical coefficients. These coefficients are then
chosen to reproduce the non minimal remainder of $A_2$. This procedure
gives\footnote{Note that the
sign of the fourth term of $B_2$ is incorrect in
\cite{Chan:1986jq}.}
\begin{eqnarray}
B_0 &=& \big\langle N \big\rangle_z  \,,
\nonumber \\ 
B_1 &=& 
\big\langle
-\frac{1}{2} N_\mu^2
+\frac{1}{6} \R N
\big\rangle_z \,,
\label{eq:5.2}
\\
B_2 &=&
\big\langle
-N_\mu^2 N_\nu^2
+ \frac{1}{2} (N_\mu N_\nu)^2
+ (NN_{\mu\mu})^2
\nonumber \\ &&
+ 2N N_\mu N_\nu N Z_{\mu\nu}
+ \frac{1}{2} (N Z_{\mu\nu} N)^2
\nonumber \\ &&
-\frac{1}{12}  \R N_\mu^2
-\frac{2}{3} \Rc_{\mu\nu} N N_\mu N N_\nu
+\frac{1}{30} \R_{\mu\mu} N
\nonumber \\ &&
+\frac{1}{72} \R^2 N
-\frac{1}{180} \Rc_{\mu\nu}^2 N
+\frac{1}{180} R_{\mu\nu\alpha\beta}^2 N
\big\rangle_z 
\,,
\nonumber
\end{eqnarray}
where we use the shorthand notation
\begin{equation}
\langle 
~\rangle_z := \int_\Gamma\frac{dz}{2\pi i}e^z (
~
) \,.
\end{equation}
$B_2$ in $X$-form, (\ref{eq:4.1}), has been obtained from this
$N$-form.  Eqs. (\ref{eq:5.2}) are the extension of Chan's formulas to
curved space-time.  Once again, to this order, the only non minimal
term surviving is $-\frac{2}{3} \Rc_{\mu\nu} N N_\mu N N_\nu$ if
$X^\prime$ is used throughout.

Using (\ref{eq:2.22}), $A_n$ in $N$-form is easily obtained from
$B_n$. This gives\footnote{Let us note that the minimal parts of $A_2$
in $X$-form and in $N$-form are different:
$$
A_{2,N}^m= A_{2,X}^m - 8 I_{3,1,3} \Rc_{\mu\nu}X_\mu X_\nu \,.
$$}
\begin{eqnarray}
A_0 &=& \big\langle N \big\rangle_z  \,,
\nonumber \\ 
A_1 &=& 
\big\langle
N N_{\mu\mu} N
+\frac{1}{6} \R N
\big\rangle_z \,,
\nonumber \\ 
A_2&=&
\big\langle  
  2 N^2 N_{\mu \mu \nu \nu} N^2 
\nonumber \\ &&
+  4 N N_{\mu} N_{\mu \nu} N_{\nu} N 
+  2 N N_{\mu} N_{\nu \nu} N_{\mu} N 
\nonumber \\ &&
+  2 N N_{\mu}^2 N_{\nu \nu} N 
+  2 N N_{\mu \mu} N_{\nu}^2 N 
\nonumber \\ &&
+  4 N N_{\mu} N_{\mu \nu \nu} N^2 
+  4 N^2 N_{\mu \nu \nu} N_{\mu} N 
\nonumber \\ &&
+ 2 N^2 N_{\mu \mu}^2 N 
+  2 N N_{\mu \mu} N N_{\nu \nu} N 
+  2 N N_{\mu \mu}^2 N^2 
\nonumber \\ &&
+  4 N^2 Z_{\mu \nu} N_{\mu} N_{\nu} N 
+ 4 N N_{\mu} N_{\nu} Z_{\mu \nu} N^2 
\nonumber \\ &&
+  4 N N_{\mu} N Z_{\mu \nu} N_{\nu} N 
\nonumber \\ &&
+  2 N^2 Z_{\mu \mu \nu} N N_{\nu} N 
+ 2 N N_{\mu} N Z_{\nu \mu \nu} N^2 
\nonumber \\ &&
+ 2 N^2 Z_{\mu \nu} N Z_{\mu \nu} N^2
\nonumber \\ &&
+\frac{1}{6} \R N N_{\mu\mu} N
+\frac{1}{4} \R_\mu N N_\mu N
\nonumber \\ &&
+\frac{1}{30} \R_{\mu\mu} N
+\frac{4}{3} \Rc_{\mu\nu} N^2 N_{\mu\nu} N^2
\nonumber \\ &&
+\frac{2}{3} \Rc_{\mu\nu} N N_\mu N_\nu N
+\frac{4}{3} \Rc_{\mu\nu} N N_\mu N N_\nu N
\nonumber \\ &&
+\frac{1}{72} \R^2 N
-\frac{1}{180} \Rc_{\mu\nu}^2 N
+\frac{1}{180} R_{\mu\nu\alpha\beta}^2 N
\big\rangle_z 
\,. 
\end{eqnarray}

As noted, the existence of a Chan's form for the coefficients was not
completely obvious a priori in the curved case. The fact that this
Chan's form exists suggests that perhaps Chan's method could find a
suitable extension in the case of curved space.

\section{The effective action}
\label{sec:6}

After functional integration, the effective action of a complex
bosonic field is given by $-\Tr(\log K)$. This can be related to the
heat kernel by
\begin{equation}
-\Tr\log K= \int_0^\infty\frac{d\tau}{\tau} \Tr (e^{\tau K})
\,.
\end{equation}

Upon restoring $\tau$ in the expressions, a contribution
$I_{r_1,\ldots,r_n}{\cal O}$ in $B_n$ picks up a factor
$\tau^{\gamma+1-\rho-d/2}$, where $2\gamma$ is the mass dimension of
the operator ${\cal O}$ and $\rho=\sum_{i=1}^n r_i$, (e.g.,
$\tau^{4+1-4-d/2}$ for $I_{2,2}\R X_\mu X_\mu$ in $B_2$). After
carrying out the integral over $\tau$, the integral over $z$ in
$I_{r_1,\ldots,r_n}$ can be traded by a momentum integral. This gives
the replacement rule for going from the heat kernel to the effective
action
\begin{equation}
\frac{1}{(4\pi)^{d/2}} I_{r_1,\ldots,r_n}{\cal O}
\to
I^{\rho-\gamma}_{r_1,\ldots,r_n}{\cal O}
\end{equation}
where we have defined
\begin{equation}
I^k_{r_1,\ldots,r_n} :=
\frac{\Gamma(d/2)}{\Gamma(k+d/2)}
\int\frac{d^dq}{(2\pi)^d}(q^2)^k
N_1^{r_1} \cdots N_n^{r_n}\,,
\end{equation}
with $N=(q^2-X)^{-1}$. (Note that $I^k_{r_1,\ldots,r_n}$ depends also
on $d$.) The contributions to the effective action may be ultraviolet
and infrared divergent. Dimensional regularization applies here. These
integrals have been computed in \cite{Salcedo:2000hp} using minimal
subtraction.

The replacement rule gives for the effective action expanded in
derivatives
\begin{equation}
-\tr\log K= \int d^dx\sqrt{g}\sum_{n=0}^\infty \tr\, W_n
\end{equation}
with
\begin{eqnarray}
W_0 &=& I_1^1 \,, 
\nonumber \\ 
W_1 &=& -\frac{1}{2} I_{2,2}^1 X_\mu X_\mu
+\frac{1}{6} I_1^0\R \,, 
\nonumber \\ 
W_2 &=&
(-I_{2,2,2,2}^2+4I_{3,1,3,1}^2)X_\mu X_\mu X_\nu X_\nu 
\nonumber \\ && 
+ \frac{1}{2}I_{2,2,2,2}^2 X_\mu X_\nu X_\mu X_\nu 
+ 4I_{3,1,3}^2 X_\mu X_\mu X_{\nu\nu} 
\nonumber \\ && 
+ I_{3,3}^2 X_{\mu\mu}X_{\nu\nu} +
2I_{2,2,2}^2 X_\mu X_\nu Z_{\mu\nu} 
+ \frac{1}{2}I_{2,2}^2 Z_{\mu\nu}Z_{\mu\nu} 
\nonumber \\ && 
-\frac{1}{12} I_{2,2}^0 \R X_\mu X_\mu 
- \frac{2}{3} I_{3,3}^2 \Rc_{\mu\nu} X_\mu X_\nu
\nonumber \\ && 
+ \frac{1}{30} I_2^0 \R_{\mu\mu} 
+ \frac{1}{72} I_2^0\R^2
\nonumber \\ && 
- \frac{1}{180} I_2^0 \Rc_{\mu\nu}^2 
+ \frac{1}{180} I_2^0 R_{\mu\nu\alpha\beta}^2 
\,.
\end{eqnarray}

\section{Summary}
\label{sec:7}

We have derived, for the first time, expressions valid to all orders
in $X$ and to four covariant derivatives for the diagonal matrix
elements and also for the trace of the heat kernel operator in a
Riemannian curved manifold. The expressions presented check previously
available results for the so-call minimal case \cite{Gusynin:1990bu}
and, when reexpanded in powers of $X$, they reproduce the known HMDS
coefficients to four derivatives. We also extend Chan's formula,
originally derived for the effective action, to include curvature. As
in the flat case, the expressions in Chan's form are remarkably simple
also in the curved case. This simplicity suggests a direct calculation
of the energy-momentum tensor taking a variation of the effective
action with respect to the metric. Such a calculation has not been
addressed here. The method of covariant symbols allows to consider
more general coordinate connections, including torsion. This would be
of interest in the derivation of the Lorentz group generators since
the coordinate connection couples to them. A virtue of our formulas is
that they are equally simple for scalar wavefunctions and for tensor
ones, without redefining the gauge connection to include the parallel
transport of the new tetrad fields indices. This is achieved through a
consistent use of the operators $Z_{\mu_1\cdots\mu_n}$ which are
defined so that they are multiplicative. Remarkably the formulas
obtained are independent of the tensor representation of the
wavefunction even for the coefficients $B_n(x)$. This is because the
general formulas for the trace cyclic property and integration by
parts can also be written in a tensor representation independent way
(see Appendix \ref{app:B}).

\begin{acknowledgments}
This work is supported in part by funds provided by the Spanish DGI
and FEDER funds with grant FIS2005-00810, Junta de Andaluc{\'\i}a
grants FQM225-05, FQM481 and P06-FQM-01735 and EU Integrated
Infrastructure Initiative Hadron Physics Project contract
RII3-CT-2004-506078.
\end{acknowledgments}

\appendix

\section{Covariant symbols}
\label{app:A}

Let us give some details on the calculation of $A_n$ using covariant
symbols.  Fuller details on the use of covariant symbols with
derivative expansions can be found in \cite{Salcedo:2000hx} for flat
space-time and \cite{Salcedo:2006pv} for curved space-time.  Using the
expansions (\ref{eq:3.5}) in (\ref{eq:3.4}), reexpanding the result
and keeping terms with at most two covariant derivatives, gives
\begin{eqnarray}
\langle x|e^K|x\rangle 
 &=&\int_\Gamma\frac{dz}{2\pi i}
\frac{1}{\sqrt{g}} \int \frac{d^dp}{(2\pi)^d} e^z
\nonumber \\ &&\times
\Big[
N- N X_\alpha\,\partial^\alpha N
+ N X_\alpha\,\partial^\alpha N X_\beta\,\partial^\beta N
\nonumber \\ &&
+N \big(
\frac{1}{2!}X_{\alpha\beta}\,\partial_\alpha\partial_\beta
+\frac{1}{6}\R
+ Z_{\alpha \beta} \, p_\alpha \partial_\beta
\nonumber \\ &&
-\frac{1}{3}\Rc_{\alpha \beta} \, p_\alpha \partial_\beta
+\frac{1}{3} R_{\alpha \mu \beta \nu} \,
p_\alpha p_\beta \partial_\mu\partial_\nu
\big) N
\nonumber \\ &&
+{\cal O}(\nabla^4)
\Big] \,,
\end{eqnarray}
with $N=(z-p_\mu^2-X)^{-1}$. (Let us warn that we are using a purely
imaginary $p_\mu$, to avoid the proliferation of $i$'s in the
formulas.) The derivatives with respect to $p_\mu$ are then carried
out (using $\partial_\mu N=2p_\mu N^2$). After that, the shift $z\to
z+p_\mu^2$ allows to isolate the $p_\mu$ dependence in integrals of
the type $\int d^dp \,e^{p_\mu^2}p_{\mu_1}\cdots p_{\mu_n}$ which are
easily evaluated. These steps produce
\begin{eqnarray}
\langle x|e^K|x\rangle 
 &=&\frac{1}{(4\pi)^{d/2}}\int_\Gamma\frac{dz}{2\pi i} e^z
\nonumber \\ &&\times
\Big[
N 
+ N X_{\mu\mu} N^2
- 2 N X_{\mu\mu} N^3
\nonumber \\ &&
+ 2 N X_\mu N X_\mu N^2
- 4 N X_\mu N X_\mu N^3
\nonumber \\ &&
- 2 N X_\mu N^2  X_\mu N^2
+ \frac{1}{3}N^3\R
\nonumber \\ &&
+{\cal O}(\nabla^4)
\Big] \,.
\end{eqnarray}
with $N=(z-X)^{-1}$. This expression immediately translates into those
in (\ref{eq:3.6}) and (\ref{eq:3.7}).

\section{Trace and integration by parts}
\label{app:B}

In this work (and in particular in this appendix), all formulas hold
for arbitrary tensor representations of the wavefunctions $\psi(x)$ on
which the Klein-Gordon operator acts. The space of world tensors of
rank $r$ is spanned by $e^{\mu_1}_{a_1}\cdots e^{\mu_r}_{a_r}$ where
$e^{\mu}_{a}(x)$ is a local basis of the space-time tangent space.

The trace $\tr$ in (\ref{eq:2.19}) refers to gauge indices and to
world indices, $\tr=\tr_{\text{gauge}}\tr_{\text{world}}$. We need to
consider only multiplicative operators $\hat{\cal O}$ (that is,
$\hat{\cal O}\psi$ at $x$ does not depend on $\psi$ at $x^\prime\not=
x$). All operators in $A_n$ and $B_n$ are multiplicative. In this
case, the trace on world indices in the representation of tensors of
rank $r$ will be
\begin{equation}
\tr_{\text{world}}(\hat{\cal O}) =
e_{\mu_1}^{a_1}\cdots e_{\mu_r}^{a_r}
(\hat{\cal O}
e^{\mu_1}_{a_1}\cdots e^{\mu_r}_{a_r})
\end{equation}
where $e_\mu^a$ is the dual basis, $e_\mu^a e^\mu_b=\delta_{ab}$.
Note that $\tr_{\text{world}}(\hat{\cal O})$ may only be non vanishing
when $\hat{\cal O}$ is itself a world scalar, that is, it maps tensors
of rank $r$ to tensors of rank $r$.  As illustration, in the space of
tensors of rank $r$, an easy calculation yields
\begin{eqnarray}
\tr\left(\frac{1}{12}Z_{\mu\nu}^2 \right) &=& 
d^r\tr_{\text{gauge}}\left(\frac{1}{12}\Omega_{\mu\nu}^2 \right) 
\nonumber \\ &&
-r d^{r-1} n_g\frac{1}{12}R_{\mu\nu\alpha\beta}^2 \,,
\end{eqnarray}
$n_g$ being the dimension of the gauge representation.  

The use of the trace cyclic property and integration by parts requires
some care. These properties work as usual when the operators involved
are (i) multiplicative and (ii) {\em they do not act on world
indices}. (For a scalar operator $\hat{\cal O}$, this means that the
component $\mu$ of $\hat{\cal O}$ applied to $e_a(x)$ does not depend
on $e_a^\nu(x)$, for $\nu\not=\mu$.) Instances of such operators are
$X$, $X_\mu$ and $\Omega_{\mu\nu}$. In the notation of
\cite{Salcedo:2006pv}, these are the operators in the class ${\cal
C}(\underline{\nabla},\underline{Z})$.
On the other hand, modifications occur in the trace cyclic property
and integration by parts when multiplicative operators acting on world
indices are involved (class ${\cal C}(\underline{\nabla})$).  An
instance of this is $Z_{\mu\nu}$, since $(Z_{\mu\nu}e_a)^\lambda=
R_{\mu\nu\lambda\sigma}e_a^\sigma + \Omega_{\mu\nu}e_a^\lambda$.
(However, $[\nabla_\mu,A]$ and $[Z_{\mu\nu},A]\in {\cal
C}(\underline{\nabla},\underline{Z})$ provided $A\in {\cal
C}(\underline{\nabla},\underline{Z})$.)

To write down the correct relations, let us define
$Z^R_{\mu_1\cdots\mu_n}$ as the curvature parts of
$Z_{\mu_1\cdots\mu_n}$, that is, obtained by dropping $\omega_\mu$ in
the covariant derivative. In particular,
\begin{equation}
Z_{\mu\nu}=Z^R_{\mu\nu}+\Omega_{\mu\nu} \,.
\end{equation}
Furthermore, let us introduce the shorthand notation
\begin{equation}
\big\langle
\hat{\cal O}
\big\rangle
:=
\int d^dx\, \sqrt{g}\,\tr(\hat{\cal O}) \,.
\end{equation}

The two following useful properties are easily established
\begin{eqnarray}
\big\langle
A Z^R_{\mu\nu}
\big\rangle &=& 0
\,, 
\quad
A\in {\cal C}(\underline{\nabla},\underline{Z})
\nonumber \\ 
\big\langle A Z^R_{\alpha\mu\nu} \big\rangle &=&
\big\langle
-\frac{1}{2} R_{\sigma\sigma\alpha\mu\nu} A
\big\rangle,
\quad
A\in {\cal C}(\underline{\nabla},\underline{Z})
\end{eqnarray}
which hold for arbitrary multiplicative operators $A$ not acting on
world indices.

Using these properties, one can prove the following relations for
arbitrary operators $A,B,\ldots$, in ${\cal
C}(\underline{\nabla},\underline{Z})$:
\begin{widetext}
\begin{eqnarray}
\big\langle
[\nabla_\mu,A]B
\big\rangle
&=&
\big\langle
-A[\nabla_\mu,B]
\big\rangle
\,,
\\ 
\big\langle
[\nabla_\alpha,A]B Z_{\mu\nu} C
\big\rangle &=&
\big\langle
-A(B_\alpha Z_{\mu\nu} C + B Z_{\alpha\mu\nu} C
+B Z_{\mu\nu} C_\alpha
+R_{\sigma\alpha\mu\nu}BC_\sigma
+\frac{1}{2}R_{\sigma\sigma\alpha\mu\nu}BC
)
\big\rangle 
\,,
\\ 
\big\langle
AB Z_{\mu\nu} C
\big\rangle &=&
\big\langle
B Z_{\mu\nu} C A - B C [Z^R_{\mu\nu},A]
\big\rangle 
\,,
\label{eq:B10}
\\ 
\big\langle
AB Z_{\alpha\mu\nu} C
\big\rangle &=&
\big\langle
B Z_{\alpha\mu\nu} C A - B C [Z^R_{\alpha\mu\nu},A]
\big\rangle 
\,,
\\ 
\big\langle
AB Z_{\mu\nu} C Z_{\alpha\beta} D
\big\rangle &=&
\big\langle
B Z_{\mu\nu} C Z_{\alpha\beta} DA
- B Z_{\mu\nu} C D [Z^R_{\alpha\beta},A ]
- B C Z_{\alpha\beta} D[ Z^R_{\mu\nu},A ]
+  B C D [Z^R_{\alpha\beta}, [Z^R_{\mu\nu},A ]]
\big\rangle 
\,,
\\ 
\big\langle
Z_{\mu\nu} A Z_{\alpha\beta} B
\big\rangle &=&
\big\langle
A Z_{\alpha\beta} B Z_{\mu\nu}
- R_{\alpha\beta\mu\sigma} A B Z_{\sigma\nu}
- R_{\alpha\beta\nu\sigma} A B Z_{\mu\sigma}
\big\rangle 
\,.
\end{eqnarray}
\end{widetext}
The first two identities refer to integration by parts while the other
allow to apply the trace cyclic property.  The relation (\ref{eq:4.2})
is a consequence of (\ref{eq:B10}).


\end{document}